\documentstyle[12pt]{article}

  \def\pp{{\mathchoice
          {
              \kern 1pt%
              \raise 1pt
              \vbox{\hrule width5pt height0.4pt depth0pt
                    \kern -2pt
                    \hbox{\kern 2.3pt
                          \vrule width0.4pt height6pt depth0pt
                          }
                    \kern -2pt
                    \hrule width5pt height0.4pt depth0pt}%
                    \kern 1pt
           }
            {
              \kern 1pt%
              \raise 1pt
              \vbox{\hrule width4.3pt height0.4pt depth0pt
                    \kern -1.8pt
                    \hbox{\kern 1.95pt
                          \vrule width0.4pt height5.4pt depth0pt
                          }
                    \kern -1.8pt
                    \hrule width4.3pt height0.4pt depth0pt}%
                    \kern 1pt
            }
            {
              \kern 0.5pt%
              \raise 1pt
              \vbox{\hrule width4.0pt height0.3pt depth0pt
                    \kern -1.9pt  
                    \hbox{\kern 1.85pt
                          \vrule width0.3pt height5.7pt depth0pt
                          }
                    \kern -1.9pt
                    \hrule width4.0pt height0.3pt depth0pt}%
                    \kern 0.5pt
            }
            {
              \kern 0.5pt%
              \raise 1pt
              \vbox{\hrule width3.6pt height0.3pt depth0pt
                    \kern -1.5pt
                    \hbox{\kern 1.65pt
                          \vrule width0.3pt height4.5pt depth0pt
                          }
                    \kern -1.5pt
                    \hrule width3.6pt height0.3pt depth0pt}%
                    \kern 0.5pt
            }
        }}

  \def\mm{{\mathchoice
                       {
                             \kern 1pt
               \raise 1pt    \vbox{\hrule width5pt height0.4pt depth0pt
                                  \kern 2pt
                                  \hrule width5pt height0.4pt depth0pt}
                             \kern 1pt}
                       {
                            \kern 1pt
               \raise 1pt \vbox{\hrule width4.3pt height0.4pt depth0pt
                                  \kern 1.8pt
                                  \hrule width4.3pt height0.4pt depth0pt}
                             \kern 1pt}
                       {
                            \kern 0.5pt
               \raise 1pt
                            \vbox{\hrule width4.0pt height0.3pt depth0pt
                                  \kern 1.9pt
                                  \hrule width4.0pt height0.3pt depth0pt}
                            \kern 1pt}
                       {
                           \kern 0.5pt
             \raise 1pt  \vbox{\hrule width3.6pt height0.3pt depth0pt
                                  \kern 1.5pt
                                  \hrule width3.6pt height0.3pt depth0pt}
                           \kern 0.5pt}
                       }}

\catcode`@=11
\def\un#1{\relax\ifmmode\@@underline#1\else
        $\@@underline{\hbox{#1}}$\relax\fi}
\catcode`@=12

\let\du=\du                     

\def\a{\alpha}
\def\b{\beta}
\def\c{\chi}
\def\d{\delta}

\def\f{\phi}
\def\g{\gamma}
\def\h{\eta}

\def\k{\kappa}
\def\l{\lambda}
\def\m{\mu}
\def\n{\nu}
\def\o{\omega}

\def\q{\theta}
\def\r{\rho}
\def\s{\sigma}

\def\z{\zeta}

\def\F{\Phi}

\def\L{\Lambda}
\def\O{\Omega}

\def\ve{\varepsilon}

\def\cd{{\cal D}}
\def\ce{{\cal E}}
\def\cf{{\cal F}}

\def\cw{{\cal W}}

\def\cy{{\cal Y}}

\def\bo{{\raise-.3ex\hbox{\large$\Box$}}}               
\def\pa{\partial}                                       
\def\de{\nabla}                                         
\def\pr{\prod}                                          
\def\TH{{\raise.2ex\hbox{$\displaystyle \bigodot$}\mskip-4.7mu \llap H \;}}
\def\face{{\raise.2ex\hbox{$\displaystyle \bigodot$}\mskip-2.2mu \llap {$\ddot
        \smile$}}}                                      

\def\sp#1{{}^{#1}}                              
\def\leftrightarrowfill{$\mathsurround=0pt \mathord\leftarrow \mkern-6mu
        \cleaders\hbox{$\mkern-2mu \mathord- \mkern-2mu$}\hfill
        \mkern-6mu \mathord\rightarrow$}
\def\dvec#1{\vbox{\ialign{##\crcr
        \leftrightarrowfill\crcr\noalign{\kern-1pt\nointerlineskip}
        $\hfil\displaystyle{#1}\hfil$\crcr}}}           
\def\dt#1{{\buildrel {\hbox{\LARGE .}} \over {#1}}}     

\def\frac#1#2{{\textstyle{#1\over\vphantom2\smash{\raise.20ex
        \hbox{$\scriptstyle{#2}$}}}}}                   
\def\sfrac#1#2{{\vphantom1\smash{\lower.5ex\hbox{\small$#1$}}\over
        \vphantom1\smash{\raise.4ex\hbox{\small$#2$}}}} 
\def\bfrac#1#2{{\vphantom1\smash{\lower.5ex\hbox{$#1$}}\over
        \vphantom1\smash{\raise.3ex\hbox{$#2$}}}}       
\def\afrac#1#2{{\vphantom1\smash{\lower.5ex\hbox{$#1$}}\over#2}}    

\def\[{\lfloor{\hskip 0.35pt}\!\!\!\lceil}
\def\]{\rfloor{\hskip 0.35pt}\!\!\!\rceil}
\def\Lag{{\cal L}}
\def\du#1#2{_{#1}{}^{#2}}

\def\dud#1#2#3{_{#1}{}^{#2}{}_{#3}}

\def\fracm#1#2{\hbox{\large{${\frac{{#1}}{{#2}}}$}}}
\def\ha{{\fracmm12}}
\def\tr{{\rm tr}}

\def\un{\underline}
\def\fracmm#1#2{{{#1}\over{#2}}}

\def\low#1{{\raise -3pt\hbox{${\hskip 0.75pt}\!_{#1}$}}}

\def\Dot#1{\buildrel{_{_{\hskip 0.01in}\bullet}}\over{#1}}
\def\dt#1{\Dot{#1}}

\newskip\humongous \humongous=0pt plus 1000pt minus 1000pt

\newif\ifdtup

\def\ref#1{$\sp{#1)}$}

\def\pl#1#2#3{Phys.~Lett.~{\bf {#1}B} (19{#2}) #3}
\def\np#1#2#3{Nucl.~Phys.~{\bf B{#1}} (19{#2}) #3}
\def\prl#1#2#3{Phys.~Rev.~Lett.~{\bf #1} (19{#2}) #3}
\def\pr#1#2#3{Phys.~Rev.~{\bf D{#1}} (19{#2}) #3}
\def\cqg#1#2#3{Class.~and Quantum Grav.~{\bf {#1}} (19{#2}) #3}

\def\mpl#1#2#3{Mod.~Phys.~Lett.~{\bf A{#1}} (19{#2}) #3}

\parskip=\medskipamount                 
\thicklines                         

\begin{document}

\thispagestyle{empty}

{\hbox to\hsize{
\vbox{\noindent UMDEPP 01--051   \hfill  hep-th/0104223 \\ 
                KL ~TH~ 01/03 \hfill     April 2001     }}}

\noindent
\vskip1.3cm
\begin{center}

{\Large\bf 4D, $N \, =\, 1$ Born-Infeld Supergravity}

\vglue.3in

S. James Gates, Jr.
\footnote{Supported in part by NSF grant \# PHY--98--02551}

{\it Department of Physics\\
     University of Maryland\\
     College Park, MD 20742, USA}
\vglue.1in
{\sl gates@physics.umd.edu}
\vglue.1in
and
\vglue.1in
Sergei V. Ketov 
\footnote{Supported in part by the `Deutsche Forschungsgemeinschaft'}

{\it Department of Theoretical Physics\\
     Erwin Schr\"odinger Strasse \\
     University of Kaiserslautern}\\
{\it 67653 Kaiserslautern, Germany}
\vglue.1in
{\sl ketov@physik.uni-kl.de}

\end{center}
\vglue.2in
\begin{center}
{\Large\bf Abstract}
\end{center}

We propose the 4D, $N \,=\,$ 1 supergravitational analogues (avatars) 
of the 4D, $N \,=\,$ 1  supersymmetric Born-Infeld action in four
dimensions for the first  time, by using superspace. In particular, 
a new Born-Infeld type generalization of the Weyl supergravity action 
is given.  A natural  new Born-Infeld type  generalization of the 
Einstein supergravity is found as well.  We also brielfy  discuss a
construction of the four-dimensional Born-Infeld-Einstein  supergravity
from the AdS supergravity in five dimensions, which seems to be very
natural in our approach.

\newpage

\section{Introduction}

~~~~The {\it Born-Infeld} (BI) electrodynamics \cite{bi} is the
non-linear (gauge- and Poincar\'e -invariant) generalization of 
the Maxwell electrodynamics. The BI theory shares with the Maxwell
theory  electric-magnetic self-duality \cite{bid} and physical
propagation  (no shock waves) \cite{caus}, which are quite non-trivial 
in the non-linear  case. Similarly to the Maxwell Lagrangian, 
the BI Lagrangian is independent  upon the spacetime derivatives 
of the Maxwell field strength,
$$ 
\Lag\low{\rm BI}(F) = \fracmm{1}{b^2} \left\{ 
 1- \sqrt{ -\det (\h_{\m\n}+bF_{\m\n} )}\right\} ~,
\eqno(1.1)$$
where $F_{\m\n}=\pa_{\m}A_{\n}-\pa_{\n}A_{\m}$, $\m,\n=0,1,2,3$, 
and $b$ is the dimensional coupling constant. The BI action is also 
known to be the low-energy bosonic part of the (gauge-fixed) effective action 
of a D3-brane filling in four spacetime dimensions (see 
ref.~\cite{tse} for a recent review). The corresponding $N \,=\,$ 1 
supersymmetric abelian BI action in four dimensions is the
Goldstone-Maxwell  action associated with {\it Partial} (1/2) 
{\it {Spontaneous Supersymmetry Breaking}} (PSSB) $N \,=\,$ 2 to $N \,=\,$ 1, 
whose Goldstone fields belong to a (Maxwell) vector supermultiplet 
with respect to unbroken $N \,=\,$ 1 supersymmetry 
\cite{bg,tr}.~\footnote{See also refs.~\cite{dp,srf} for an earlier 
construction of the $N\,=\,1$ supersymmetric BI actions.}  The 
similar $N \,=\,$ 2 supersymmetric abelian BI action \cite{sk2} is the 
relevant (low-energy) part of the effective worldvolume action 
of a D3-brane in six dimensions \cite{sk2e}, since the $N \,=\,$ 2 BI 
action \cite{sk2} is the most relevant part of the $N \,=\,$ 2 
Goldstone-Maxwell
action associated with the PSSB $N \,=\,$ 4 to $N \,=\,$ 2, modulo terms with
 spacetime
derivatives of the $N \,=\,$ 2 Maxwell  superfield strength
\cite{sk2,sk2e,kuz,bik}. 

It is of considerable interest to construct possible {\it gravitational} 
analogues of the BI action (see, e.g., refs.~\cite{deg,gha} for earlier 
discussions without supersymmetry). A supersymmetric BI action possesses 
more physically important features when compared to its purely bosonic 
BI part (e.g. PSSB), while supersymmetry also implies more constraints 
on a BI-type  (non-linear in the curvature) supergravity when compared 
to the purely bosonic theory. This is the main idea of this paper, which
is based on manifest local  $N \,=\,$ 1 supersymmetry as the sole construction
tool. Our investigation may be  considered as part of the more ambitious
programm of summing up gravitational corrections in string theory. To 
the best of our knowledge, no reasonable proposal for a BI-supergravity
action was ever made. Further constraints  like PSSB, ghost freedom and
duality on the top of manifest local $N \,=\,$ 1 supersymmetry will be 
considered elsewhere.

Supersymmetry does not necessarily prefer the standard determinantal 
form of the BI action in eq.~(1.1). Moreover, eq.~(1.1) is not even 
the most compact (and, hence, most elegant and simple) form of the
BI  theory! The bosonic variable having the most natural $N \,=\,$ 1
supersymmetric extention (with linearly realized $N \,=\,$ 1 supersymmetry in
superspace)  is given by \cite{bg}
$$
\o=\a +i\b~,\quad{\rm where}\quad 
\a = \fracmm{1}{4}F^{\m\n}F_{\m\n}\equiv  \fracmm{1}{4}F^2\quad 
{\rm and}\quad \b = \fracmm{1}{4}F^{\m\n}\tilde{F}_{\m\n}\equiv  
\fracmm{1}{4}F\tilde{F}~,
\eqno(1.2)$$
and $\tilde{F}_{\m\n}$ is the dual tensor, $\tilde{F}_{\m\n}=\ha\ve_{
\m \n\l\r}F^{\l\r}$. The BI Lagrangian (1.1) can be rewritten to the 
form $(b=1)$ 
$$ 
\Lag\low{\rm BI}(\o,\bar{\o})= 1- \sqrt{ 1+ (\o+\bar{\o})
+\frac{1}{4}(\o-\bar{\o})^2}~~,
\eqno(1.3)$$
or, equivalently, 
$$  
\Lag\low{\rm BI}(\o,\bar{\o})= \Lag\low{\rm ~free}+ \Lag\low{\rm 
~int.} \equiv  -\,\fracm{1}{2}\left(\o+\bar{\o}\right)
+\o\bar{\o}\cy(\o,\bar{\o})~,
\eqno(1.4)$$
with the structure function 
$$
\cy(\o,\bar{\o})\equiv \fracmm{1}{1+\fracmm{1}{2}(\o+\bar{\o})+
\sqrt{ 1+ (\o+\bar{\o})+\fracmm{1}{4}(\o-\bar{\o})^2}}~~.
\eqno(1.5)$$
The remarkably simple equivalent form of the BI action \cite{bg,tr},
$$  
\Lag\low{\rm BI}(\o,\bar{\o})= -\,\fracmm{1}{2}\left(\c+\bar{\c}
\right) =-{\rm Re}\,\c~,
\eqno(1.6)$$   
arises as a solution to the non-linear constraint
$$ 
\c =-\,\fracm{1}{2}\c\bar{\c}+\o~.
\eqno(1.7)$$
This {\it Non-Linear Sigma-Model} (NLSM) form of the BI action is 
quite natural from the viewpoint of PSSB \cite{tr,book}. Indeed, 
to spontaneously break any rigid symmetry, one may start with a 
free action that is invariant under the linearly realized symmetry, 
and then impose an invariant non-linear  constraint that gives rise 
to the NLSM whose solutions break the symmetry.

Unlike the Maxwell action minimally coupled to gravity,
$$
S_{\rm M}=-\,\fracm{1}{4}\int d^4x\sqrt{-g}\,g^{\m\l}g^{\n\r}
F_{\m\n}F_{\l\r}~,
\eqno(1.8)$$ 
which is invariant under the local Weyl transformations, 
$$ 
g_{\m\n}\to e^{2\l(x)} g_{\m\n}\,\quad A_{\m} \to  A_{\m}~,
\eqno(1.9)$$
the BI action minimally coupled to gravity is obviously not 
Weyl-invariant.  Nevertheless, the BI action can be made Weyl-invariant 
by using a conformal  compensator $\f(x)$ with the Weyl 
transformation law
$$ 
\f \to  e^{-\l(x)}\f~. 
\eqno(1.10)$$
The modified BI action 
$$ 
S =\int d^4x \left\{ \sqrt{-\det(\f^2 g_{\m\n})} -     
   \sqrt{-\det(\f^2g_{\m\n} +F_{\m\n})} \right\} 
\eqno(1.11)$$
is obviously invariant under the transformations (1.9) and (1.10). 
Equation (1.11) can be further generalized to \cite{deg}
$$ 
S_{\rm DG} =\int d^4x\left\{ \sqrt{-\det ( \f^2 g_{\m\n})} -    
\sqrt{-\det(\f^2g_{\m\n} +\f^{-2}D_{\m}\f D_{\n}\f +F_{\m\n})}
\right\}~,
 \eqno(1.12)$$
which is also Weyl-invariant when using the Weyl-covariant derivative 
$D_{\m}=\pa_{\m}\f +A_{\m}$ and the abelian gauge transformation law 
$A_{\m} \to  A_{\m} +\pa_{\m}\l$. The use of Weyl (conformal) 
compensators is the standard tool in supergravity \cite{gbook}. 

There are many ways to add a non-minimal coupling to gravity in the 
BI action as well as to build new `pure BI gravity' actions by using 
curvature-dependent terms. For example, if one insists on the 
determinantal form of the bosonic BI gravity, one can simply 
substitute the metric $g_{\m\n}$ under the square root of the 
determinant by $(g_{\m\n}+\k^2R_{\m\n})$, where $R_{\m\n}$ is the 
Ricci tensor of the metric $g_{\m\n}$ and $\k$ is a constant of
dimension of length,
$$ 
S =\fracmm{1}{\k^4}\int d^4x\left\{ \sqrt{-\det ( g_{\m\n})} -    
\sqrt{-\det\left(g_{\m\n} +\k^2R_{\m\n}\right)}~\right\}~. 
\eqno(1.13)$$
The expansion of this action in powers of $\k^2$ yields the 
Einstein-Hilbert action as the leading contribution. The equations 
of motion for the action (1.13) are satisfied by any Ricci-flat 
metric, e.g. by the Schwarzschild black hole metric.  Hence, any 
BI-type action merely depending upon the Ricci tensor is not going 
to remove the black hole singularity at the origin (no taming). In 
other words, the Weyl tensor should also enter the  BI-type gravity
action, and we now need a reasonable proposal for it. Of  course, 
one could simply extend the determinantal prescription by using 
the most general substitution under the square root of the 
determinant, {\it viz.}
$$
g_{\m\n}\to g_{\m\n}+\k^2R_{\m\n}+\z X_{\m\n}~,
\eqno(1.14)$$ 
where $X_{\m\n}$ is {\it any} tensor quadratic or higher in the full 
curvature, and $\z$ is yet another dimensional constant \cite{deg}. 
Unfortunately, all these prescriptions do not have natural 
supersymmetric extensions and, therefore, they are ignored in what
follows. 

The gravitational analogues of the Maxwell variables in eq.~(1.2) are 
given by
$$ 
\a_{\rm G}= R^{\m\n\l\r}R_{\m\n\l\r} \quad {\rm and}\quad 
\b_{\rm G}= R^{\m\n\l\r}\tilde{R}_{\m\n\l\r}~,
\eqno(1.15)$$
where $R_{\m\n\l\r}$ is the full (Riemann-Christoffel) curvature tensor,
$\tilde{R}_{\m\n\l\r}$ is the dual (in curved spacetime) curvature, 
while all indices are raised and lowered by the use of $g_{\m\n}$ and 
its inverse $g^{\m\n}$  as usual. The BI-NLSM prescription (1.7) gives
rise to the non-linear (in the curvature) Born-Infeld-Gravity Lagrangian
$$  \Lag_{\rm BI2G}=-\sqrt{-g}\,{\rm Re}\,\c_{\rm G}~,\quad  
 \c_{\rm G} =-\,\fracm{1}{2}\c_{\rm G}\bar{\c}_{\rm G}+\o_{\rm G}~,
\quad {\rm and}\quad \o_{\rm G}=\a_{\rm G}+i\b_{\rm G}~,\eqno(1.16)$$
which appears to be the BI-type extension of the quadratically-generated 
gravity with $\Lag_{\rm 2G}=-\sqrt{-g}\,R^2_{\m\n\l\r}$. Requiring the
existence of a  locally $N \,=\,$ 1 supersymmetric extension rules out the
bosonic BI action (1.16),  though it becomes possible after replacing 
the curvature tensor by the Weyl  tensor above (sect.~3). 

The action (1.16) does not have the Einstein-Hilbert (linear in the 
curvature) term. Of course, in the bosonic theory this drawback can 
be easily corrected, e.g. by combining the two prescriptions above,
$$ 
\Lag_{\rm BIG}=\left[ \sqrt{-\det(g_{\m\n})} -\sqrt{-\det(g_{\m\n}+\k^2
R_{\m\n})} \right]\left(\k^{-4}+{\rm Re}\c_{\rm G}\right)~,
\eqno(1.17)$$
in order to get the Einstein-Hilbert term as the leading contribution. 
Of course, this is rather artificial and, in fact, it does not have a 
supersymmetric extension too. Some possible resolutions to this problem 
are given in the next sections. 

To the end of this section we give a brief comment about ghosts.  As 
is well-known, the Gauss-Bonnet density $\sqrt{-g}\b_{\rm G}$ or 
$$ 
E = \sqrt{-g}\left[R^2_{\m\n\l\r}-4R^2_{\m\n}+R^2\right] 
\eqno(1.18)$$
is a total derivative in four dimensions, so that any 
quadratically-generated gravity is actually proportional to a linear
conbination of R$^2_{\m\n}$ and $R^2$, which always results in the
presence of ghosts in its free part.   The ghost-freedom to this 
order can be obtained by arranging the ghost-free combination, $
\sqrt{-g}\left(c_1-\fracm{1}{2}R\right)+c_2E$ with some  constants
$(c_1,c_2)$ as the leading contribution (modulo curvature-cubed  
terms).  It is worth mentioning, however, that the standard BI action
(1.1) is actually ghost-free because of its special non-polynomial 
(or non-perturbative) structure. The quartic terms in the BI action 
are proportional to the  Euler-Heisenberg term, $(F^2)^2+(F\tilde{F})^2$, 
whereas the following terms are not ghost-free at any given (finite)
 order. It would be interesting to check whether any of the actions (1.16) 
or (1.17) is ghost-free or not.

As will be shown in the next sections, manifest local $N \,=\,$ 1 
supersymmetry 
leads to certain restrictions on the bosonic BI gravity, while the
ghost-freedom is not automatically resolved by introducing supersymmetry
alone.  We believe that it is very natural to search for a
supergravitational BI action by using a reformulation of gravity as 
a non-abelian gauge theory.  This then implies first a construction 
of the {\it Non-abelian} Born-Infeld (NBI) theory. The  $N \,=\,$ 1 
supersymmetric
NBI action, proposed by one of the authors in  ref.~\cite{kno} (see
sect.~2), is going to be used in sect.~3 for a  construction of the BI
supergravity action.

\section{$N \,=\,$ 1 supersymmetric BI and NBI actions}

~~~~In this section we briefly recall the $N \,=\,$ 1 supersymmetric abelian BI 
action in superspace \cite{bg,tr} and its $N \,=\,$ 1  {\it Non-abelian} (NBI) 
generalization in four dimensions \cite{kno}. 

The BI action (1.1) in the form (1.6) is most convenient for 
supersymmetrization in superspace. One replaces the abelian bosonic 
field strength $F_{\m\n}$ by the abelian $N \,=\,$ 1 chiral spinor superfield
strength $W_{\a}$ obeying the $N \,=\,$ 1 superspace Bianchi identities 
$$ 
\bar{D}_{\dt{\a}}W\low{\a}=0,\qquad D^{\a}W_{\a}-\bar{D}_{\dt{\a}}
\bar{W}^{\dt{\a}}=0~,\qquad \a=1,2~,
\eqno(2.1)$$ 
while the superextension of $\o$ is simply given by $W^2$. The $N \,=\,$ 1 
manifestly supersymmetric abelian BI action \cite{bg} in the NLSM 
form reads \cite{tr}
$$ 
S_{\rm 1BI}=\int d^4xd^2\q\,X +{\rm h.c.},
\eqno(2.2)$$
where the $N \,=\,$ 1 chiral superfield Lagrangian $X$ obeys the non-linear 
constraint 
$$ 
X = \ha X\bar{D}^2\bar{X}+\ha W^{\a}W_{\a}~~.
\eqno(2.3)$$
The iterative solution to eq.~(2.3) gives rise to the superfield 
action \cite{bg}
$$ 
S_{\rm 1BI} = \frac{1}{2}\left(\int d^4xd^2\q\,W^2+{\rm h.c.}\right) 
+ \int d^4xd^4\q\,\cy(\frac{1}{2} D^2W^2,\frac{1}{2}
\bar{D}^2\bar{W}^2)W^2\bar{W}^2 
\eqno(2.4)$$ 
with {\it the same} structure function (1.5) of the bosonic BI theory.
 
It is worth mentioning that the NLSM form (2.2) of the $N \,=\,$ 1 BI action is 
also most useful in proving its invariance under the second (non-linearly
realized and spontaneously broken) supersymmetry with rigid spinor
parameter $\h^{\a}$ \cite{bg,tr},
$$ 
\d_2 X=\h^{\a}W_{\a}~,\quad \d_2W_{\a}=\h_{\a}\left(1-\ha\bar{D}^2
\bar{X} \right)+i\bar{\h}^{\dt{\a}}\pa_{\a\dt{\a}}X~~,
\eqno(2.5)$$
and its $N \,=\,$ 1 supersymmetric electric-magnetic self-duality as well. The 
latter amounts to a verification of the non-local constraint \cite{kuz}
$$ \int d^4x d^2\q (W^2+M^2)= \int d^4x d^2\bar{\q} (\bar{W}^2+\bar{M}^2
)~, \quad{\rm where}\quad  \fracmm{i}{2} M_{\a}=\fracmm{\d S_{\rm
1BI}}{\d W^{\a}}~~.
\eqno(2.6)$$

It is not difficult to put the $N \,=\,$ 1 supersymmetric BI action (2.2) into 
the $N \,=\,$ 1 superconformal form by inserting the conformal compensator ($N \,=\,$ 1
chiral superfield) $\F$ into the non-linear constraint (2.3) as follows
\cite{kuz}:
$$ 
X = \fracmm{X}{2\F^2}\bar{D}^2\left(\fracmm{\bar{X}}{\bar{\F}^2}
\right) +\ha W^{\a}W_{\a}~~.
\eqno(2.7)$$ 
Equation (2.2) is recovered from eq.~(2.7) in the gauge $\F=1$.

The simple structure of the $N \,=\,$ 1 supersymmetric abelian BI action (2.2) 
dictated by the Gaussian non-linear constraint (2.3) allows us to easily
construct its non-abelian (NBI) generalization \cite{kno} that may also 
be relevant for the effective description of the D3-brane clusters (i.e.
the D3-branes on the top of each other). 

The $N \,=\,$ 1 {\it Super-Yang-Mills} (SYM) theory in $N \,=\,$ 1 superspace is 
defined by the standard off-shell constraints \cite{gbook}:
$$ 
\{\de\low{\a},\de\low{\b} \}= \{ \bar{\de}_{\dt{\a}},\bar{\de}_{
\dt{\b}} \}=0~, \quad \{ \de\low{\a},\bar{\de}_{\dt{\b}}
\}=-2i\de_{\a\dt{\b}}~~,
$$
$$ 
\[ \de\low{\a},\de_{\b\dt{\b}} \] =2i\ve\low{\a\b}\hat{
\bar{W}}_{\dt{\b}}~~, \quad \[ \bar{\de}_{\dt{\a}},\de_{\b\dt{\b}} \]
=2i\ve_{\dt{\a}\dt{\b}} \hat{W}\low{\b}~~,
\eqno(2.8)$$
in terms of the $N \,=\,$ 1 covariantly-chiral (Lie algebra-valued) gauge 
superfield strength $\hat{W}_{\a}=\hat{W}_{\a}^at_a$ obeying the Bianchi 
identities~\footnote{The Lie algebra generators $t_a$ obey the relations
$\[t_a,t_b\]=f_{abc}t_c$ and $\tr(t_at_b)=-2\d_{ab}$.} 
$$ 
\bar{\de}_{\dt{\a}}\hat{W}\low{\a}=0~,\quad \bar{\de}_{\dt{\a}}\hat{
\bar{W}}{}^{\dt{\a}}=\de^{\a}\hat{W}\low{\a}~~.
\eqno(2.9) $$

The natural $N \,=\,$ 1 supersymmetric NBI action is \cite{kno} 
$$  
S\low{\rm NBI} = \int d^4xd^2\q\,\tr~\hat{\F} + {\rm h.c.}~,
\eqno(2.10) $$
whose $N \,=\,$ 1 covarianlty chiral Lagrangian $\hat{\F}$ is subject to 
the `minimal' non-abelian generalization of the abelian non-linear
constraint (2.3),
$$  
\hat{\F} = \frac{1}{2} \hat{\F}\bar{\de}^2\hat{\bar{\F}}+\frac{1}{2} 
 \hat{W}^2~~.
\eqno(2.11)$$ 

The leading contribution to the NBI action (2.10) is the standard $N \,=\,$ 1 
SYM action in superspace \cite{gbook},
$$ 
S_{\rm SYM}= \fracm{1}{2} \int d^4xd^2\q\,\tr~\hat{W}^2  + {\rm h.c.}
\eqno(2.12)$$
The next (NBI) correction in the Yang-Mills sector (in components) 
\cite{kno},
$$
\fracm{1}{4}\tr\left[ (F^2)^2 + (F\tilde{F})^2\right]~,
\eqno(2.13)$$ 
appears to be the non-abelian version of the Euler-Heisenberg term
$\fracm{1}{4}\left[ (F^2)^2 + (F\tilde{F})^2\right]$ which is present 
in the BI Lagrangian (1.1) as the leading $(F^4)$ correction to the
Maxwell term.

The $N \,=\,$ 2 super BI and NBI actions \cite{sk2,sk2e,kuz,bik,kno,za} 
have similar features.      

\section{Born-Infeld supergravity}

~~~~As is well-known, $N \,=\,$ 1 supergravity in four dimensions is most
naturally described in curved superspace $z^M=(x^m,\q^{\m}, \bar{
\q}_{\dt{\m}})$, $m=0,1,2,3$ and $\m=1,2$, where we now have to 
distinguish between curved $(M)$ and flat $(A)$ indices related
by a supervielbein $E\du{A}{M}$ and its inverse $E\du{M}{A}$ with 
$E={\rm Ber}(E\du{A}{M})\neq 0$  \cite{gbook}.  The supervielbein 
$E\du{A}{M}$ and a superconnection $\O_A$ are most conveniently 
described by (super)  one-forms,
$$ 
E_A=E\du{A}{M}(z)\pa_M\quad {\rm and}\quad \O=dz^M\O_M(z)
= E^A\O_A~,
\eqno(3.1)$$
where $\O_A$ take their values in the Lorentz algebra,
$$ 
\O_A=\ha\O\du{A}{bc}(z)M_{bc}=\O\du{A}{\b\g}M_{\b\g}+
\O\du{A}{\dt{\b}\dt{\g}}\bar{M}_{\dt{\b}\dt{\g}}~,
\eqno(3.2)$$
and $M\low{bc}~\sim~(M\low{\b\g},M_{\dt{\b}\dt{\g}})$ are the Lorentz 
generators.  The curved superspace covariant derivatives
$$ 
\cd_A=\left(\cd_a,\cd_{\a},\bar{\cd}^{\dt{\a}}\right)=E_A+\O_A
\eqno(3.3)$$ 
obey the algebra
$$
\[ \cd_A,\cd_B\}=T\du{AB}{C}\cd_C+{\cal R}_{AB}~,
\eqno(3.4)$$
where the supertorsion $T\du{AB}{C}$ and the (Lorentz algebra-valued) 
supercurvature ${\cal R}_{AB}=\ha {\cal R}\du{AB}{cd}M_{cd}={\cal
R}\du{AB}{\b\g}M_{\b\g} + {\cal R}\du{AB}{\dt{\b} \dt{\g}} \bar{M}_{
\dt{\b}\dt{\g}}$ have been introduced. 

The universal (conformal) supergravity constraints can be divided into 
three sets. The first set of the constraints is needed for the existence
of  chiral superfields in curved superspace --- these are the so-called 
representation-preserving constraints:
$$ 
T\du{\dt{\a}\dt{\b}}{c}=T\du{\dt{\a}\dt{\b}}{\g}=T\du{\a\b}{c}=
T\du{\a\b}{\dt{\g}}=0~.
\eqno(3.5a)$$
The second set of the constraints is needed to solve the vector covariant 
derivative in terms of the spinor ones --- these are the so-called 
conventional constraints of type-I:
$$ T\du{\a\dt{\b}}{\g}= T\du{\a\dt{\b}}{\dt{\g}}={\cal R}\du{\a\dt
{\b }}{cd} =0~,\quad T\du{\a\dt{\b}}{c}=-2i(\s^c)_{a\dt{\b}}~.
\eqno(3.5b)$$ 
The third set of the constraints is used to determine the spinor 
superconnections in terms of the spinor supervielbeins --- these are the 
so-called  conventional constraints of type-II:
$$  
T\du{\a\b}{\g}= T\du{\dt{\a}\dt{\b}}{\dt{\g}}= T\dud{\a,\b(\dt{\b},}
{\b}{\dt{\g})}=T\du{\dt{\a},(\b\dt{\b},\g)}{\dt{\b}}=0~.
\eqno(3.5c)$$

The so-called {\it minimal} $N \,=\,$ 1 supergravity arises by adding extra
constraints,
$$ 
T_{\a,\b\dt{\b},\g\dt{\g}}=\ve\low{\a\b}\ve_{\dt{\b}\dt{\g}}T_{\g}
=0 \quad {\rm and}\quad T_{\dt{\a},\b\dt{\b},\g \dt{\g}}=
\ve_{\dt{\a}\dt{\b}} \ve\low{\b\g}\bar{T}_{\dt{\g}}=0~.\eqno(3.5d)$$

Taken together, equations (3.5) are just the standard (Wess-Zumino) 
$N \,=\,$ 1 
supergravity constraints \cite{wzu}. As a result of the constraints and 
the Bianchi identities, all the superfield components of the supertorsion
and the  supercurvature appear to be merely dependent upon three
(constrained) supertorsion tensors: the complex (covariantly) chiral
scalar superfield ${\cal R}$, the real vector superfield $G_a$ and the
complex (covariantly)  chiral superfield $W_{\a\b\g}$ that is totally
symmetric with respect to its  spinor indices \cite{gbook}. The bosonic
superfield ${\cal R}$ has an  auxiliary  complex scalar $B$ as the
leading component, while it also contains the spacetime scalar curvature
as another bosonic field component. Similarly,  the bosonic vector
superfield  $G_a$ has the spacetime Ricci curvature amongst its field 
components. The fermionic superfield $W_{\a\b\g}$ has the gravitino
field strength as its leading component, while it also contains the 
spacetime Weyl tensor $C_{\a\b\g\d}$ (totally symmetric on its spinor
indices) as the fermionic field component.

The $N \,=\,$ 1 (Einstein) supergravity action \cite{gbook}
$$ 
S_{\rm SG} = -\,\fracmm{3}{\k^2}\int d^8z E^{-1} 
\eqno(3.6)$$
is just the supervolume of curved $N \,=\,$ 1 superspace. Here $\k$ is the
gravitational coupling constant of dimension of length.  A chiral 
local denisty also exists in curved $N \,=\,$ 1 superspace \cite{gbook},
$$ 
\ce=-\fracmm{1}{4}{\cal R}^{-1}(\bar{\cd}^2-4{\cal
R})E^{-1}~.
\eqno(3.7)$$  
The simple `covariantizing' rules in the chiral $N \,=\,$ 1 superspace are 
given by 
$$ 
d^4x d^2\q\to d^4x d^2\q\,\ce \quad{\rm and}\quad
 \bar{D}^2 \to (\bar{\cd}^2-4{\cal R})~. 
\eqno(3.8)$$

The BI-type non-linear superfield constraint (2.11) can be considered 
as the powerful tool converting {\it any} fundamental (input) chiral
superfield  Lagrangian $(\hat{W}^2)$ into the corresponding BI-type
chiral Lagrangian $(\hat{\F})$ in superspace. A natural candidate for 
the BI supergravity action just arises along these lines.  Indeed, the
supergravitational analogue of  the (covariantly chiral) $N \,=\,$ 1 SYM spinor
superfield strength $W_{\a}^at_a$ is  given by the super-Weyl curvature 
tensor $W_{\a\b\g}M^{\b\g}$ that is covariantly chiral in $N \,=\,$ 1 curved 
superspace of the $N \,=\,$ 1 minimal supergravity. This essentially amounts 
to replacing the  Yang-Mills gauge group in the $N \,=\,$ 1 NBI action (sect.~2)
by the Lorentz group.  The Weyl supergravity action \cite{gbook}
$$ S_{\rm W}= \int  d^4x d^2\q\,\ce\,\tr\,W^2 +{\rm h.c.}\eqno(3.9)$$ 
can then be extended to the corresponding {\it Born-Infeld-Weyl} (BIW) 
supergravity action 
$$ 
S_{\rm BIW}=\int   d^4x d^2\q\ce\,\tr\,\cf +{\rm h.c.}~,
\eqno(3.10)$$
whose covariantly chiral (Lorentz algebra-valued) Lagrangian $\cf$ is 
a solution to the non-linear superfield constraint
$$ 
\cf= \fracm{1}{2} \cf(\bar{\cd}^2-4{\cal R})\bar{\cf}+W^2~.
\eqno(3.11)$$
It is worth noticing that the superfield ${\cal R}$ also enters the 
action (3.11). 

The subleading correction to the Weyl supergravity action in the BIW 
theory (3.10) is given by
$$ 
S_{\rm BR}=\fracm{1}{2}\int  d^4x d^4\q E^{-1} \, {W^2}\low{\a\b\g}
\bar{W}^2_{\dt{\a}\dt{\b}\dt{\g}}~,
\eqno(3.12)$$ 
whose purely bosonic (gravitational) part is proportional to the square 
of the {\it Bel-Robinson} (BR) tensor \cite{br}
$$ 
T_{mnpq}= R_{mspt}R\du{n}{s}{}\du{q}{t}+R_{msqt}R\du{n}{s}{}\du{p}{t}
-\fracm{1}{2}g_{mn}R_{prst}R\du{q}{rst}~~.
\eqno(3.13)$$
In {\it four} dimensions, the BR tensor (3.13) can be identically 
rewritten to
$$  
T_{mnpq}=R_{mspt}R\du{n}{s}{}\du{q}{t} + \tilde{R}_{mspt}\tilde{R}
\du{n}{s}{}\du{q}{t}~,
\eqno(3.14)$$ 
where  $\tilde{R}_{mspt}$ is the dual curvature. Moreover, in four
dimensions the BR tensor is known to be symmetric in all four indices 
and pairwise traceless \cite{br}. The BR tensor squared, $T^2_{mnpq}\,$,
should therefore  be considered as the gravitational analogue of the
subleading  (Euler-Heisenberg) term, $(F^2)^2+(F\tilde{F})^2$, in the
abelian BI action (1.1).
 
Of course, unlike the Weyl supergraviy action (3.9), the BIW action 
(3.10) is no longer invariant under the super-Weyl transformations (this
is similar to the BI action vs. the Maxwell action). However, the
conformal invariance can  be easily restored by introducing a conformal
compensator (the covariantly  chiral $N \,=\,$ 1 scalar superfield) of Weyl
weight $(-1)$, as in eq.~(2.7).  

Though the BIW action naively seems to be the most obvious gravitational 
analogue of the NBI action, even the leading terms of the BIW action 
contain terms with higher derivatives. Unlike the gauge theory that is
quadratic in the field strength, the Einstein action is linear in the
curvature, in components. The $N \,=\,$ 1 supergravity action does not contain
the supercurvature at all (i.e. of the zeroth order). Hence, it is
worthwhile to investigate how the Einstein supergravity may appear in 
our approach.
 
A {\it Born-Infeld-Einstein} (BIE) supergravity can be generated in 
several ways, either from the quadratic action in AdS five dimensions
(sect.~4)  by using the abelian BI machinery in the form (2.2) and 
(2.3), or simply by inserting supertensors into the action (3.6). In
fact, {\it any} full superspace action containing a supercurvature (even
linearly) gives rise to the terms that are non-linear in the component
curvature. As an example, let's consider the most `economical' superfield
Lagrangian whose superspace structure resembles the component Einstein
action with a cosmological term,
$$ 
S_{\rm BIE} =\int d^8z E^{-1} (\L+{\cal R}) +{\rm h.c.}~,
\eqno(3.15)$$
where $\L$ is a non-vanishing constant. In components, this very simple 
local $N \,=\,$ 1 superinvariant gives rise to the following bosonic terms: 
$$ 
S_{\rm bos.} = -\,\fracm{1}{9} \int d^4x\sqrt{-g}(R+\fracm{1}{3}
B\bar{B}) (2\L+B+\bar{B})~,
\eqno(3.16)$$
where the auxiliary complex scalar field $B$ is the leading component of 
${\cal R}$. The algebraic $B$-equation of motion has an obvious solution 
$$ 
B=\bar{B}= -\frac{1}{3}\L \pm \sqrt{\frac{1}{9}\L^2 -R}~.
\eqno(3.17)$$
Being inserted back into the action (3.16), this yields   
$$ 
S_{\rm bos.} = -\fracm{4}{27}\int d^4x\sqrt{-g}\left\{\fracm{4}{3}
\L R + (\frac{1}{9}\L^2-R)\left(\frac{1}{3} \L\mp\sqrt{\frac{1}{9}
\L^2-R}\right)\right\}~.
\eqno(3.18)$$
This action is already of the BI type, while it also implies taming of
the scalar curvature from above,
$$ 
R\leq (\frac{1}{3}\L)^2~.
\eqno(3.19)$$

After choosing the upper sign (minus) choice in eq.~(3.18) and adjusting 
the free parameter $\L$ as
$$ 
\L = \left( \fracmm{3}{2\k}\right)^2~, 
\eqno(3.20)$$
where $\k$ is the gravitational constant of dimension of length as usual,
the leading term (in the curvature) in the action (3.18) takes the 
standard (Einstein-Hilbert) form, $-\fracmm{1}{2\k^2}R~$.

As is well known, one of the most beautiful features of the original BI 
action is its famous taming of the Coulomb (electro-magnetic field)
self-energy of a  point-like electric charge \cite{bi}. Related to this
feature is the existence of the maximal value for the electro-magnetic
field strength. Similarly, one may expect from a BIE action that it
should remove the spacetime singularity of the Schwarzschild (or
Schwarzschild-AdS) solution (black hole) in the Einstein or AdS 
theory. This can only happen if Ricci-flat solutions are excluded, 
which is not the case for the BIE action (3.15). We may thus need a 
better BIE action that would be dependent upon the Weyl supertensor
$W_{\a\b\g}$ too,  e.g., by combining eqs.~(3.10) and (3.15).  A more
natural solution within our  approach apparently implies a generation 
of the Einstein term from some action that is quadratic in the curvatures
(sect.~4).

\section{AdS-BI supergravity}

~~~~In this section we propose yet another approach for a construction 
of gravitational and supergravitational avatars of the BI action.  This
component  approach is based on treating gravity and supergravity as 
the  geometrical theories by gauging the AdS (covering) symmetry group
$Sp(4)$ or  the AdS supergroup $OSp(1,4)$, respectively \cite{mm}. Here
we briefly outline this construction in the supergravity case.

The MacDowell-Mansouri procedure consists of the following 
steps~\cite{mm}: 
\begin{itemize}
\item take the spacetime supersymmetry $OSp(1,4)$ as the gauge group, 
and introduce the connection one-forms $h^A=h^A_{\m}dx^{\m}$ and the
corresponding curvature two-forms $\cw^A$ a l\'a Yang-Mills, $\cw
\du{\m\n}{A}=\pa_{\m}h^A_{\n}-\pa_{\n}h^A_{\m}+h_{\m}^Bh_{\n}^C
f\du{BC}{A}$, where $f\du{BC}{A}$ are the structure constants of
$OSp(1,4)$, \item define the invariant action
$$ 
S_{\rm AdS}= \int \cw^A\wedge \cw^B Q_{AB}~,
\eqno(4.1)$$
where $Q_{AB}=\{\ve_{abcd},const.(C\g_5)_{\a\b}\}$, $\ve_{abcd}$ is the 
Levi-Civita symbol, $C_{\a\g}$ is a charge conjugation matrix, and 
$const.\neq 0$,
\item
varying the action (4.1) with respect to the connections associated 
with the Lorentz generators allows one to express the Lorentz connections 
$h_{\m}^{[ab]}$ in terms of the remaining gauge fields (the vierbein
$h^a_{\m}$ and a gravitino $h^{\a}_{\m}$).  This also converts the 
first-order action (4.1) into the second-order action and ensures its
invariance under supersymmetry transformations, \item a decomposition 
of the action $ S_{\rm AdS}$ with respect to the irreducible
representations of $OSp(1,4)$ results in a sum of the {\it topological}
Euler-Poincar\'e characteristic of the four-dimensional Lorentz base
manifold (spacetime), the {\it Einstein} gravity term, and a {\it
cosmological} term, together with their fully supersymmetric completion.
\end{itemize}

The MacDowell-Mansouri approach \cite{mm} for a construction of the 
Einstein supergravity puts both gravity and supergravity theories on 
equal footing, by generating both of them from the most basic gauge 
field theory action  quadratic in the curvature. Therefore, their 
approach is perfectly suitable  for our purpose of generating a BIE 
action via the BI equations (1.6) and  (1.7), with the MacDowell-Mansouri
density $\ve^{\m\n\l\r}R\du{\m\n}{A} R\du{\l\r}{B}Q_{AB}$ 
as the input $(\o)$.

\section{Conclusion}

~~~~In this paper we proposed the new BIW, BIE and AdS-BI actions with 
 manifest local $N \,=\,$ 1 supersymmetry.
Our  construction is entirely based on the non-abelian and locally
supersymmetric  generalization of the non-linear constraint (1.7)
governing the structure of the BI action (1.1). This mechanism is related
to spontaneous partial supersymmetry breaking, while it seems to be
similar to the renormalization procedure converting the bare (input)
coupling constant into the `running'  (effective) coupling constant in a
renormalizable quantum field theory. Of course, it would be interesting
to know whether some of our actions survive  tests of causual propagation
and/or positivity of energy. We are also going to investigate a possible
connection between our actions and the effective  actions for D-branes.

We conclude with two comments. 

The bosonic non-abelian BI action is well-known to suffer from the
non-abelian ambiguities \cite{tse}.  Our NBI prescription does not have
these ambiguities since the iterative solution to  the NBI constraint
(2.11) also implies definitive ordering of the non-abelian  quantities.
This equally applies to our BI supergravity actions. 

It is not difficult to generalize our $N \,=\,$ 1 BI supergravity actions 
to $N \,=\,$ 2 and $N \,=\,$ 4 extened BI supergravity too. 
The off-shell $N \,=\,$ 2 supercurvature, supertorsion and chiral density 
superfields are well known in the standard curved $N \,=\,$ 2 superspace 
\cite{mull}. For example, the Weyl tensor is hiding in the $N \,=\,$ 2 bosonic
 (covariantly chiral) superfield $W_{\a\b}$ that is symmetric  on its spinor 
indices, being the $N \,=\,$ 2 analogue to the $N \,=\,$ 1 Weyl superfield
 $W_{\a\b\g}$ (see, e.g., ref.~\cite{kgn}). Similarly, there exists a complex
  chiral scalar $N \,=\,$ 4 superfield $W$ containing the Weyl
tensor in the $N \,=\,$ 4  superfield supergravity \cite{n4sg}.

\newpage

\section*{Acknowledgements}

We are grateful to Stanley Deser and Gary Gibbons for discussions.
\vglue.2in

\end{document}